\documentclass{article}
\usepackage{spconf,amsmath,graphicx}
\usepackage[table]{xcolor}
\usepackage[pagebackref=true,breaklinks=true,colorlinks,bookmarks=false,citecolor=blue,linkcolor=blue]{hyperref}

\usepackage{multirow}

\newcommand{\etal}{\textit{et al.}}
\graphicspath{{grf/}}
\title{Image-based EEG classification of Brain Responses to Song Recordings}
\name{Adolfo G.~Ramirez-Aristizabal$^{1}$, Mohammad K.~Ebrahimpour$^{2}$, Christopher T.~ Kello$^{1}$}

\address{University of California,~Merced$^{1}$, Ericsson Inc.$^{2}$}
\begin{document}
%
\maketitle
\begin{abstract}
 Classifying EEG responses to naturalistic acoustic stimuli is of theoretical and practical importance, but standard approaches are limited by processing individual channels separately on very short sound segments (a few seconds or less). Recent developments have shown classification for music stimuli ($\sim$ 2 mins) by extracting spectral components from EEG and using convolutional neural networks (CNNs). 
This paper proposes an efficient method to map raw EEG signals to individual songs listened for end-to-end classification. EEG channels are treated as a dimension of a [$\textit{Channel} \times \textit{Sample}$] image tile, and images are classified using CNNs. Our experimental results (~88.7\%) compete with state-of-the-art methods (85.0\%), yet our classification task is more challenging by processing longer stimuli that were similar to each other in perceptual quality, and were unfamiliar to participants. We also adopt a transfer learning scheme using a pre-trained ResNet-50, confirming the effectiveness of transfer learning despite image domains being unrelated from each other.      
\end{abstract}
\section{Introduction}
\label{s:intro}
Brain Computer Interface (BCI) research seeks to interpret information retained in brain responses that relates to perceived stimuli. Traditional BCI approaches have leveraged correlated behaviors measured through brain responses, such as modeling the relationship between ocular directionality or mouse tracking and cortical activity~\cite{1,2}. When feeding these data to deep learning networks, classification of specific actions across varying contextual scenes becomes strong e.g., leveraging pupil dilation with cortical activity to classify click actions when navigating the internet~\cite{3}.
 \begin{figure}[t]
\includegraphics[width = 8.75cm, height = 4.75cm]{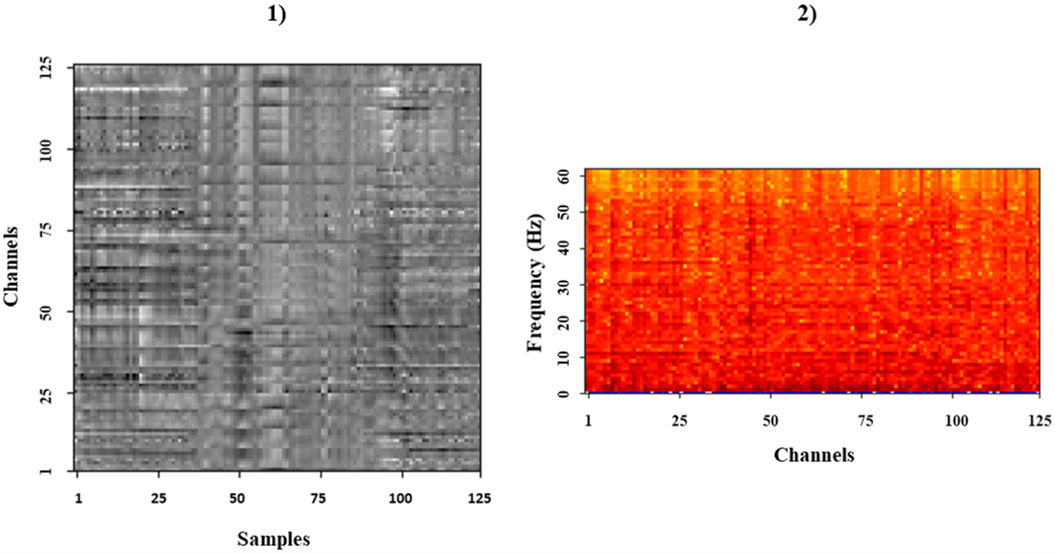} \\
\caption{\small 1) the grayscale and 2) the PSD input representations of participant 1 at the 100th second of song 1. 2) one second PSD at 125 hz produced up to 63 Hz frequency components.}
\label{f:name}
\end{figure}

 Consequentially, this leads to interest in researching methods that could interpret the passive responses to complex, naturalistic stimuli which open the possibility of including populations with limited motor capabilities. An example of this in the image-stimuli domain has shown the ability to both classify and reconstruct image categories that participants were passively exhibited. Spampinato \etal~\cite{4} used 10 image classes from ImageNet~\cite{5} and randomly presented examples of $0.5$ second presentation length. The electroencephalogram (EEG) responses to these stimuli were then fed to a recurrent neural network (RNN).

Methods focused on only processing cortical activity have fallen under the neural decoding umbrella, with a similar goal of relating external behaviors to the internal cortical activity. One of the primary endeavors in neural decoding research is to model temporal correlations of stimuli to their corresponding brain response via regression or classification-based decoders~\cite{6}. Common practices with neural spike trains and EEG recordings have included using RNN frameworks to capture temporal transitions over time. EEG recordings show high temporal resolution but a low spatial resolution trade-off~\cite{7}. Here we focus on the comparison of EEG studies related to processing temporal relationships using deep learning methods.

Some approaches use Convolutional Neural Networks (CNNs) along with traditional feature extraction techniques. This usually involves taking a specified EEG channel and passing it a wavelet transform to turn it into a 2D spectral input representation~\cite{8,9}. Then the 2D spectral input is sent to the CNN for further processing. For instance, Supratak \etal~\cite{10} utilized the single-channel recordings along with 1D CNN layers as an input for a Long short-term memory (LSTM) model. Lawhern \etal proposed EEGNet, which leverages from the depthwise
\begin{table}[h]
    \centering
    \caption{\small Details of the proposed architecture. Input: The grayscale portrait of EEG signal. Output: The class labels associated to music genre. }
    \scalebox{0.9}{
    \begin{tabular}{c|c c c}

    \hline
    Layer Type & Filter Size & Input & Output\\
    \hline
    \hline
    Conv2D & $4 \times 4$ & $125 \times 125 \times 1$ & $63 \times 63 \times 32$ \\
    BatchNorm2D & - & - & - \\
    \hline
    Conv2D & $4 \times 4$ & $63 \times 63 \times 32$ & $32 \times 32 \times 64$ \\
    BatchNorm2D & - & - & - \\
    \hline
    Conv2D & $4 \times 4$ & $32 \times 32 \times 64$ & $16 \times 16 \times 128$ \\  
    BatchNorm2D & - & - & - \\
    \hline
    GAP & $16 \times 16$ & $16 \times 16 \times 128$ & $1 \times 1 \times 128$ \\
    \hline
    FC1 & - & 128 & 100 \\
    BatchNorm1D & - & - & - \\
    FC2 & - & 100 & 10 \\
    \hline
    \end{tabular}}
    \label{t:architecture}
\end{table}
and separable convolution technique~\cite{11}. This allows for the model to use the EEG channel ordering to approximate filter-bank common spatial pattern (FBCSP) and bilinear discriminate component analysis. This is successful because EEG channel ordering follow topographic relationships of how they are recorded on the scalp~\cite{12, 7}. 

Recent studies explored neural decoding concepts and applied them to acoustic EEG responses of complex stimuli. Stober \etal~\cite{13} presented participants with $100 ms$ long sinusoid rhythms based on tribal cadences and achieved $\sim$24.4\% classification performance on a total of 24 classes. Stronger performance was shown ($\sim$83.2\%) in a 3 class RNN model using spoken vowels ($\sim$0.5 sec) with featured extracted EEG inputs~\cite{14}. Another study used longer stimuli ($\sim$10 sec) of 8 varying types of vocalizations and was able to achieve $\sim$61\% performance without any feature extraction on EEG passed to DenseNet. Yu \etal~\cite{15} improved the performance to $\sim$81\% by incorporating canonical correlation analysis between DenseNet and pre-trained VGG model that extracted audio features of the experimental stimuli. Most recently, Sonawane \etal~\cite{16} improved on these approaches and showed that longer and complex stimuli (~2 mins of music) could be used to evoke EEG responses used as spectral 2D CNN inputs to classify song ID ($\sim$85.0\%).

In this paper, we investigate raw EEG input as a potentially efficient and readily available input representation. We created an EEG grayscale image defined by [\textit{Channels}, \textit{Samples}] dimensions. This approach presents state-of-the-art performance using a fraction of trainable parameters with indie/electronic song stimuli (4 mins each) that were unfamiliar to participants. Input representations are analyzed through Multidimensional Scaling (MDS) on channel order and compared with Power Spectral Density (PSD) feature extraction. Lastly, classification results are extended with a supporting dataset using pop Hindi songs as stimuli.

\section{Method}
\label{s:proposed_method}
We used the first 4 minutes of all recordings and split them into 5-second chunks. To create training and test sets from the same distribution, we balance the train and test data across time by assigning every other chunk to go either train or test at a 75/25 ratio. From there, the chunks were cut up into 1-second examples and the model presentation order was randomly shuffled. Input tensors were of [$\textit{Batch Size} \times \textit{Samples} \times \textit{Channels} \times \textit{Depth}$] dimensions. Given a 1 sec input length the input representations were now square images at depth of 1 giving input tensor dimension values of [$\textit{Batch Size} \times 125 \times 125 \times 1$].

Given the training dataset 
$\cal S$ $ = \cup_{i=1}^{n}\{ x_i,y_i\}$ drawn i.i.d. from distribution $\cal D$, we
seek to learn a model that generalizes well. In particular, consider a family of models parameterized by $w \in W \subseteq \mathbf{R}^{d}$. We define the training
set loss as follows:
\begin{equation}
    L(w) = \frac{1}{n} \sum_{i=1}^{n} \ell(w,x_i,y_i)
\end{equation}
where $x_i \in \mathbf{R}^{125 \times 125 \times 1}$ which is a grayscale portrait of the EEG recordings. Then, the input is fed to the CNN network and after a softmax layer the objective function to be minimized is as follows:
\begin{equation}
    L(w) = -\frac{1}{n} \sum_{1}^{n} y_i \text{log}(G(x))
\end{equation}
where $G(.)$ is the neural network followed by a softmax layer. A summary of the proposed architecture is illustrated in Table~\ref{t:architecture}.

Since the input is a grayscale image, we can apply 2D convolutional layers to extract features. The kernel size is fixed to $4 \times 4$ with a stride of $2$. The convolutional layers have $32,64,128$ filters, respectively. Since the final task is recognition, the network requires the most abstract representation. Therefore, we applied a Global Average Pooling (GAP)~\cite{17} on top of the last convolutional layer. Since most of the parameters in a neural networks comes in the fully connected layers (FC), GAP layer significantly reduces the parameters. The activation function on all convolution and fully connected layers (except the output layer) is fixed as linear Rectifier units and the network is initialized with He~\cite{17} initialization technique.

Finally, transfer learning methods have proven useful for improving the speed of training and asymptotic performance of deep learning models. Since the input is a 2D image of the EEG with spatial structure, we hypothesized that transfer learning is feasible, even when transfer stimuli come from naturalistic images as opposed to EEG images. Therefore, we adopted the ResNet50 pre-trained on ImageNet to our task~\cite{5}. Because the input to ResNet50 is a color image, we stacked our grayscale image three times to simulate what would be RGB channels.
\section{Experimental Results}
\label{sec:pagestyle}
\subsection{Datasets}
\label{ssec:dsets}
 We evaluated our method on two publicly available datasets: Naturalistic Music EEG Dataset – Tempo (NMED-T) and Naturalistic Music EEG Dataset – Hindi (NMED-H). In both datasets participants are recorded in a passive listening experiment~\cite{18, 19}. The NMED-T dataset is comprised of 20 participants who listened to 10 songs to their full length (4:30-5:00 mins) in a randomized order. The dataset also included behavioral rating questions (scaled 1-9) on both how familiar the participant was with each song and how enjoyable they found each song after listening. Familiarity ratings across participants were very low on average. The experimenters also selected the songs to have unique tempos from one another. Uniqueness was defined by having a different Beats Per Minute (BPM) and varying low-frequency spectral peaks. Furthermore, all songs contained vocals with one song not in English.

The NMED-T dataset included both the cleaned and unprocessed signals, for which we opted to use the cleaned signals as to not include muscle artifacts. Specifics of their pre-processing steps can be found in their report~\cite{19}. Independent Component Analysis (ICA) was used to remove muscular artifacts and ocular components were computed to reject bad channels. The recordings were done using the Electrical Geodesics Inc. (EGI) GES300 system with an EGI Net Amps 300 amplifier and Net Station 4.5.7 acquisition software at a 1 kHz sampling rate~\cite{20}. On the other hand, the NMED-H dataset included similar preprocessing except that it used Reliable Components Analysis (RCA) for channel selection. The main difference lies in the participant schema and stimuli, for which there were 12 different participants per condition and each stimulus was a full-length Hindi pop song. There was only a total of four different songs, and participants listened to the songs twice but in our model training, we only used data from the first listen. Lastly, guidance for cross-dataset compatibility has been published to better connect results~\cite{17}.

\subsection{Comparisons}
\label{ssec:comp}
 Results are generalized to random unheard examples across time and within participants. This follows the training approach of the visually evoked EEG response classification models~\cite{4}. Other EEG classification studies also do not do across participant generalization because of the high cost of acquiring a large number of new participants when one participant can give you sufficient recordings~\cite{14, 16}. 

The NMED-T dataset has a total of 10 class labels, and the NMED-H dataset has 4 classes. We also added yet another classification task for enjoyment ratings. The NMED-T dataset has enjoyment ratings scaled from 1-9. We applied the enjoyment ratings as targets for the classification of rating. However, enjoyment ratings were unequally distributed. Thus, we decided to group ratings into three broad groups: low, medium, and high enjoyment classes for a 3-class classification model. Standard sentiment classification studies also only use either binary or 3-class models \cite{21}, and our label reassignment seeks to make the task relevant while also compromising with the dataset size limitation. Furthermore, we include several validation metrics to more strictly interpret the effects of reassigning labels.   

\begin{table}[t]
\caption{\small Grand performance summary of all our models. Top panel includes models trained on raw EEG with using both the NMED-T and NMED-H datasets. Bottom panel shows results for models trained on the different input representations including EEG channel ordered randomly or by MDS.}
    \centering
\scalebox{0.95}{
\begin{tabular}{c|c|c|c|c}
\hline
 \textbf{Models}& \textbf{Accuracy\%}& \textbf{Precision\%}& \textbf{F1\%}& \textbf{Kappa\%}\\
 \hline
 NMED-T& 88.69& 88.85& 88.67& 87.43\\
 NMED-H & 97.09& 97.36& 97.05& 96.13\\
 ResNet-50 & 93.05& 93.08& 93.05& 92.28\\
 Enjoyment & 90.12& 90.15& 90.10& 83.87\\
 \hline 
 Random & 80.23& 80.68& 80.22& 78.04\\
 MDS & 83.13& 83.60& 83.08& 81.26\\
 PSD & 83.18& 84.05& 83.20& 81.31\\
 PSD-2 & 80.75& 81.81& 80.60& 78.61\\
 Res-PSD & 94.12 & 94.13 & 94.12 & 93.46 \\
 \hline
\end{tabular}}
    \label{t:ours_diff_ds}
\end{table}
\begin{table*}[t]
    \caption{\small Summary of studies that try to classify EEG responses to an ID label of complex auditory stimuli.}
    \centering
\begin{tabular}{c|c|c|c|c|c}
 \hline
 \textbf{Studies}& \textbf{Accuracy(\%)}& \textbf{Class Size}& \textbf{Stimuli Length}& \textbf{Feature Extraction}&\textbf{Stimuli Type}\\
 \hline
 Sonawane \etal 2021~\cite{16}& 84.96& 12&2 mins& Yes&Music\\
 \hline 
 Moinnereau \etal 2018~\cite{14}& 83.20& 3& 0.5 secs& Yes&Spoken Vowels\\
 \hline 
 Yu \etal 2018~\cite{15}& 61.00& 8& 10 secs& No&Vocals\\
 \hline 
 Stober \etal 2014~\cite{13}& 24.40& 24& 100 ms& No&Sinusoid Rhythms\\
 \hline 
 \rowcolor{lightgray} \textbf{Our Study}& \textbf{88.69}& 10& \textbf{4 mins}& No&Music\\
  \hline
\end{tabular}
    \label{t:comparisions}
\end{table*}

Table~\ref{t:ours_diff_ds} presents the testing results of our different classification models across several performance metrics. Cohen’s Kappa illustrates the strictest metric for all models. In this case, Cohen’s Kappa metric is measuring the agreeability between the target labels of the songs and the predicted labels from the classification. The usefulness of this metric is that it considers as a ratio both the difference in true and predicted target labels over the random chance of agreeability~\cite{22}. This is especially useful for our enjoyment rating model. It seems to pick up on our 3-tier class reassignment because it has the biggest difference (~7\%) in Cohen’s Kappa to other metrics, compared to only ~2\% for the other models. Top panel of \ref{t:ours_diff_ds} shows that all our models outperform any prior EEG classification attempts.

Recent attempts at classifying auditory EEG responses have achieved similar performance (~84\%) all while relying on feature extraction or simple and short acoustic stimuli~\cite{14, 16}. Study~\cite{16} provides a strong example for treating EEG responses as images in classification tasks. They show that when the EEG recordings are treated as 1D time-series in CNN layers, performance stays at chance. This is something we were able to verify with our initial attempts at classification using the NMED-T dataset. Their strongest model (84.96\%) is a consequence of extracting frequency components with a PSD analysis. In our simple 2D CNN model we show that we can achieve 88.69\% accuracy without normalizing the data or using spectral analysis. Table~\ref{t:comparisions} illustrates a comparative summary across relevant studies considering their best performance and details of the datasets used to achieve that performance. The most comparable study is by Sonawane \etal~\cite{16} and we focus here on how our approach expands on it.
Table~\ref{t:comparisions} shows that our model is trained on EEG responses with the longest stimuli by a large margin on the order of minutes. Other studies have also attempted to not include feature extraction steps but the stimuli length in their experiments are significantly smaller than ours (in the order of seconds)~\cite{13,15}. In comparison to~\cite{16}, the EEG responses in NMED-T were to unfamiliar stimuli, which in prior works has shown to be the harder case as classification performance drops when listeners are not familiar to the music stimuli~\cite{23}. Our model also achieves this level of performance (88.69\%), but with substantially smaller parameters (179,132 compared to 1,678,156) in~\cite{16}.

In any standard image classification task, it matters how the input images are represented. The use of data augmentation such as rotations, mirroring transforms, and noise can help regularize model. On the other hand, we also know that if we are classifying categories such as 'dogs', we want the face, legs, and tail to be in the correct spatial order. The concern with our input representation is whether the default channel order did not distort our proposed cortical portrait. Bottom panel of Table~\ref{t:ours_diff_ds} shows a summary of testing the input representation format.

\subsection{Input Representations}
\label{ssec:chans}
 Channel order was tested by training models with randomly shuffled channels and with channel groupings based on MDS as seen in Table~\ref{t:ours_diff_ds} with models 'Random' and 'MDS'. In short, the MDS analysis consisted of taking the root mean square (RMS) of the channel amplitude envelopes in the training data to create a difference matrix (pairwise Euclidean). The matrix was then projected into a 1-dimensional embedding through MDS manifold learning, which allowed for rank ordering channels based on their dissimilarity. Table~\ref{t:ours_diff_ds} shows that MDS ordered channel training had ~4\% less accuracy, while random ordered channels had a significant loss in performance ~13\%. With this we can have confidence that our proposed grayscale portrait with the default channel ordering is adhering to crucial spatial relationships for classification. The concept behind testing channel order is inspired by Saeed \etal~\cite{24} Channel Reordering Module (CHARM), which helps make sense of heterogeneity of electrode channels.
 
 Furthermore, we also evaluated the impact of raw input vs. spectral representation on the NMED-T dataset (visualized in Figure~\ref{f:name}). 
The periodogram function was passed to all our 1 second raw EEG inputs at 125 Hz with Fast-Fourier Transform (FFT) windows being the same size as the input length as well as passing the function to 2 seconds EEG inputs~\cite{25}. This was done because PSD approximation gives back frequency components up to half the maximum sampling rate of your input, and the 2 second EEG examples give us a representation with the same shape $\mathbf{R}^{125 \times 125}$ as our raw representation. This gives us confidence that differences in performance are between input representations and not model hyperparameters. In Table~\ref{t:ours_diff_ds} we can see that PSD representation from 1 second EEG has better performance than the 2 second examples as seen in models 'PSD' and 'PSD-2' respectively. The performance here is still ~5\% less than our main model trained on raw inputs, showing us that feature extraction is not needed for classification tasks. 
 \subsection{Transfer Learning}
\label{ssec:trans}
  Fine-tuning of the ResNet50 is done with raw and PSD representations. Both models in Table~\ref{t:ours_diff_ds}, 'ResNet-50' and 'Res-PSD', achieve test performance up to ~93\%. The high performance here is surprising because our input representations are significantly different than the original training images, as seen in Figure~\ref{f:name}. These results support our main point that EEG data can be effectively processed through standard computer vision methods to allow for stronger performance and more efficient models. We can see that with EEG we have two variables that configure a cortical portrait; channel topography and spectral extraction. More testing is needed to further understand what configurations are more appropriate across deep learning tasks, but with our results we show an example of high performing end-to-end classification. 
  
\begin{figure*}[t]
\includegraphics[width = 13.75cm, height = 5cm]{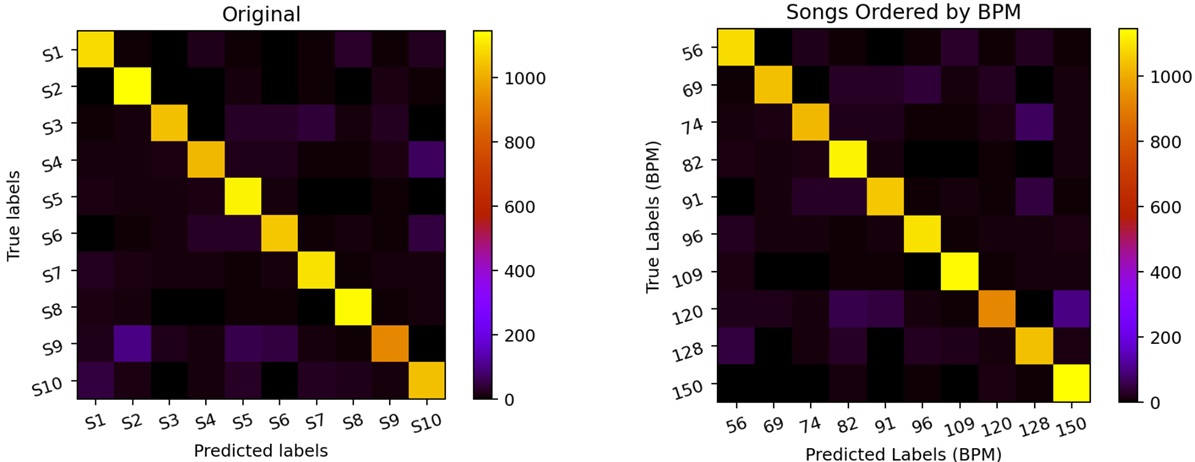} \\
\centering
\caption{\small Left, the confusion matrix for the original model. Right, results from the original model sorted by ascending BPM}
\label{f:name2}
\end{figure*}
   \subsection{Validation and Generalization}
\label{ssec:val}
  Many publicly available EEG datasets have a participant size limitation because of how expensive it is to have long recordings of various people. This primarily becomes an issue when trying to balance inputs to targets as well as how to assign train and test data. We did label reassignment on the sentiment classification task to handle the prior mentioned issue, and we separated our data into time chunks for the latter to avoid the train or test sets having too much of the beginning or end of the songs. This is also why our generalization was done using a hold out metric instead of cross-validation. In an attempt to be critical of our results, we also included precision, F1, and Cohen's Kappa as model metrics. We find that precision is higher than accuracy for all models, although not by much. We also see that F1 scores are not much lower than model accuracy but that kappa does reflect a strict interpretation of our results. We show further validation testing of our main 'NMED-T' model from Table~\ref{t:ours_diff_ds} by doing permutation tests. The first test looked at predicted performance when randomly permuting the test labels and the second test randomly permuted the model weights; in both tests performance stayed at chance 10\%.
  
  Our results are interpreted here as providing a useful method for end-to-end classification of EEG responses to music listening, but alternatively we could also be seeing these results only because the dataset was designed to be distinguishing tempos. In other words, there is a competing alternative that our model is not capturing overall acoustic features and only works because of the differences in BPM. Stober \etal~\cite{13} provides a good counterpoint to this because their results explicitly test acoustic stimuli with the exact same BPM, and test performance across their models are fairly above chance. We perform an additional analysis to address this as well, by looking at the correlation of BPM across confusions as seen in Figure~\ref{f:name2}. Specifically, we took the confusion matrix of our best performing model and organized it by ascending BPM. We saw that confusion was random and not clustered along the diagonal. Here, the average BPM difference for confusions was 31.5 BPM, whereas the chance difference was 28.95 BPM. Although the stimuli were originally picked because of their differences in tempo, some only differ with ~5 BPM, and genres fall under the similar 'indie-electronic' umbrella due to the researchers' focus on unfamiliar stimuli~\cite{19}. 
\section{Conclusion}
\label{sec:typestyle}
 Traditionally, EEG data has been likened to time-series/sequence input representations in deep learning due to the analysis conventions in Cognitive Neuroscience studies~\cite{6}. Recent relevant studies have leveraged CNNs by extracting frequency components of EEG~\cite{14, 16}. In our study, we present novel recognition results supporting our proposed method that EEG responses to full length music stimuli can achieve state-of the-art performance using the raw input without any feature extraction. We use 9 models and 2 datasets to support our method, while our main dataset used (NMED-T) is the most difficult benchmark of EEG music classification. This is because the music stimuli are the longest by an order of minutes, the participants were unfamiliar with the stimuli, and the songs overlapped significantly in terms of genre. Despite these challenges, our experiments reveal that EEG responses to music can be processed end-to-end. 
\bibliographystyle{IEEEbib}

\end{document}